# PMSR - Poor Man's Segment Routing, a minimalistic approach to Segment Routing and a Traffic Engineering use case

Stefano Salsano[(1)], Luca Veltri[(2)], Luca Davoli[(3)], Pier Luigi Ventre[(1)], Giuseppe Siracusano[(1)]

(1) Univ. of Rome Tor Vergata - (2) Univ. of Parma - (3) Univ. of Parma / Consortium GARR

*Abstract* – The current specification of the Segment Routing (SR) architecture requires enhancements to the intra-domain routing protocols (e.g. OSPF and IS-IS) so that the nodes can advertise the *Segment Identifiers* (SIDs). We propose a simpler solution called PMSR (Poor Man's Segment Routing), that does not require any enhancement to routing protocol. We compare the procedures of PMSR with traditional SR, showing that PMSR can reduce the operation and management complexity. We analyze the set of use cases in the current SR drafts and we claim that PMSR can support the large majority of them. Thanks to the drastic simplification of the Control Plane, we have been able to develop an Open Source prototype of PMSR. In the second part of the paper, we consider a Traffic Engineering use case, starting from a traditional flow assignment optimization problem which allocates *hop-by-hop* paths to flows. We propose a SR path assignment algorithm and prove that it is optimal with respect to the number of segments allocated to a flow.

*Keywords – Segment Routing, Network Architecture, Traffic Engineering, Software Defined Networking, Open Source.*

## I. INTRODUCTION

The Segment Routing (SR) architecture [1] is based on the *source routing* approach: border nodes can control the edge-to-edge routing of packets at the level of single flows by adding proper information in packet headers. This way, it offers advanced traffic steering capabilities in IP networks maintaining scalability both in the Data and Control Planes. In fact, internal nodes do not need to store any per-flow state and the traffic steering decisions have a configuration impact only on border nodes.

Segment Routing lends itself to support different applications: Virtual Private Networks (VPNs), protection/restoration, Traffic Engineering (TE), Service Function Chaining (SFC), Operation And Management (OAM). The standardization activity on the Segment Routing architecture is relatively recent. The status of the draft is mature and different independent implementations are now available. Real world deployments are ongoing, as Segment Routing has captured the interest of network providers and of "Over the Top" Providers.

On the Data Plane, the Segment Routing architecture can be implemented in different ways; in particular MPLS and IPv6 are the two Data Plane technologies that have been considered in the standardization.

Let us consider the Control Plane. In its current specification, the Segment Routing architecture [1] (section II) requires enhancements to routing protocols (e.g. [2][3]) in order to distribute the *Segment Identifiers* (SIDs). In section III we propose a minimalistic approach that does not need to explicitly distribute information among nodes and hence does not require enhancements to the routing protocols. We refer to this solution as "Poor Man's Segment Routing" (PMSR), but we claim that it can efficiently support the large majority of the use cases of traditional Segment Routing. In section III.B, we identify a set of use cases among the ones described in [4][5][6] which can be supported by the proposed solution.

In general the computation of the source routed paths and the configuration of the border nodes can be realized either in a distributed or in a centralized way. In the former case, the control logic of border nodes needs to be further enhanced. In the latter case, the Software Defined Networking (SDN) architecture [7] represents a perfect fit: a SDN approach can be used to properly configure the SR services in the border nodes, with minimal or no increase of the complexity of the border node. *The PMSR approach is in line with the SDN philosophy of removing complexity from the forwarding nodes*.

To the best of our knowledge, currently there are no Open Source implementations of the IP Control Plane extensions needed to support the traditional full-fledged SR architecture (i.e., the routing protocol enhancements). On the other hand, we have been able to fully implement the Control Plane and the Data Plane of PMSR starting from open source tools with rather limited effort [8].

In the second part of this work we focus on Traffic Engineering aspects. We start from a traditional flow assignment optimization procedure which allocates *hop-by-hop* paths to flows (section IV). Then in section V we propose a SR path assignment algorithm both for the traditional SR architecture and for the proposed PMSR. We prove that, starting from an arbitrary *hop-by-hop* path, it can evaluate the optimal SR path (i.e., the one with the minimum number of segments). We describe a simple experimental evaluation (section VI) of the algorithm, showing that its execution time is much smaller than the execution time of the flow assignment procedure.

## II. CURRENT SEGMENT ROUTING ARCHITECTURE

In the Segment Routing architecture [1] the route of a packet is enforced through an ordered list of processing/forwarding functions, called segments, that is inserted in the packet header by a border node. A segment may consist in a logical or physical element, for example a network node, a network link, or a packet filter. Each segment is identified by a Segment ID (SID). The scope of a SID can be global or local. Global SIDs are defined globally in a SR domain and are recognized by all network nodes of the domain. Instead, Local SIDs are defined locally within a node. The use of local SIDs by other nodes requires an explicit distribution mechanism or some form of centralized coordination.

Among the different types of segments described in [1], we consider Prefix segments, Node segments and IGP Adjacency segments (IGP stands for Interior Gateway Protocol). Their corresponding Segments IDs are denoted as Prefix-SIDs, Node-SIDs and Adj-SIDs.

The Prefix-SIDs represent IGP prefixes, i.e. blocks of IP addresses that are advertised, by the routing protocol, through the nodes composing the network. The routing algorithm (Shortest Path First) is used by each node to evaluate the shortest path towards the prefix and to add a corresponding entry in its routing table. With SR, a node can associate a Prefix-SID to its attached prefix and advertise it. To clarify with an example in the MPLS architecture (with absolute SIDs), a node that has the network 10.10.1.0/24 attached can associate the MPLS label 10001 as Prefix-SID and advertise this association using the routing protocol. All nodes will forward the MPLS label 10001 using the routing information available for the network 10.0.1.0/24.

A particular case of Prefix-SID is the Node-SID, which considers a /32 prefix, i.e. a single node. *"From anywhere in the network, a Node-SID enforces the ECMP-aware shortest-path forwarding of the packet towards the related node."* ([1]). In particular, the "loopback interface" address that is used to univocally refer to a router is associated to a Node-SID and advertised by each router. Even if a Node-SID is a particular type of Prefix-SID, from now on we will denote as Prefix-SIDs only the SIDs that are not Node-SIDs, i.e. those that effectively represent a range of IP addresses with a netmask different from /32.

The Node-SIDs, corresponding to the loopback interface of a node, are advertised by the node itself, while the Prefix-SIDs are advertised by the nodes that inject the routes into the IGP domain. The SID values cannot be arbitrarily chosen by the nodes, but a global coordination is needed. In fact, a SID (e.g., a MPLS label) cannot correspond to different prefixes or nodes. Quoting from [1]: *"A Prefix-SID/Node-SID is allocated […] according to a process similar to IP address allocation. Typically the Prefix-SID/Node-SID is allocated by policy by the operator (or Network Management System) and the SID very rarely changes."*. The global coordination procedure needs: i) to contact all nodes that can advertise the SIDs; ii) to configure the mapping of prefixes and loopback interface addresses to SIDs in a coordinated manner. The logical scalability of this management procedure is $O(\rho+\eta)$, where $\rho$ is the number of prefixes and $\eta$ is the number of nodes that will advertise their loopback interface. Note that the routing protocol extensions are used to automatically disseminate the mapping between SIDs and prefixes/nodes, otherwise the scalability of the configuration would become $O(\eta \cdot (\rho+\eta))$.

The third type of segment defined in [1] is the Adjacency segment. It corresponds to a unidirectional adjacency of the routing protocol, that is a specific outgoing link from a source node towards a destination node. The Adjacency segments are represented by Adj-SIDs and, usually, they are local SIDs, that can be processed only by the node that has advertised it. For example, assume that node $n$ advertises its global Node-SID $GN_n$ and one local Adj-SID $LA_{nm}$ for the outgoing interface from node $n$ to node $m$. A packet carrying the list of segments $\{GN_n, LA_{nm}\}$ will be forwarded first to node $n$, then by the node $n$ towards the node $m$. The local Adj-SID needs to be advertised by the node $n$ to all the other nodes, so that the ingress border node that evaluates SR path can include it in the segment list, but this has no impact on the routing state of the crossed nodes. It is also possible to advertise an Adjacency segment as a global segment, in the example above a global Adj-SID $GA_{nm}$ can be advertised by node $n$. The segment list to obtain the same behavior will be reduced to a single segment $\{GA_{nm}\}$, but the routing state of all nodes of the network should be dynamically updated following the distribution of the global Adj-SID. In fact, all network nodes should be capable to process the SID $GA_{nm}$, by forwarding the packet towards node $n$, while the node $n$ will be the only one that will forward the packet on its outgoing interface toward $m$. Global Adj-SIDs greatly increase the amount of routing state that needs to be maintained by nodes.

The Adj-SIDs are interesting for Traffic Engineering purposes because they allow to map an arbitrary path, composed by a sequence of links, into a list of segments. Using only Node-SIDs in SR paths, it is not possible to use links that are not chosen by the IGP protocol, such as a backup link with high assigned cost. In fact, Node-SIDs always forward packets on paths selected by the IGP protocol.

In the MPLS SR Data Plane the use of indexes has been proposed for SIDs: the MPLS label, that represents a segment is generated by combining the index value with the information related to the sets of MPLS labels made available by a given node for SR (called Segment Routing Global Block - SRGB). This approach requires the distribution of the SRGB information through extensions to the routing protocols. As mentioned in [5] *"Several operators have indicated that they would deploy the SR technology in this way: with a single consistent SRGB across all the nodes. They motivated their choice based on operational simplicity..."*. We also rule out the possibility of having different SRGBs advertised by the nodes and we only use "absolute" SIDs.

### III. POOR MAN'S SEGMENT ROUTING (PMSR)

In PMSR, we want to avoid the distribution of SIDs (Segment IDs) by the SR nodes, as it implies significant extensions to the routing protocols and to the routing daemons implementing the protocols. For this reason, we only use global segments types whose SIDs can be automatically generated by each node in a distributed fashion, with no need of explicit advertising (and no extensions to routing protocols). The automatic generation avoids the need of node management procedures for SID assignment. We advocate that a significant coverage of the SR use cases can be achieved by only using global segments that can be automatically generated.

In case of Node-SIDs, it is relatively easy to define an automatic mapping between the IP addresses of the node loopback interface and the SID. For the IPv6 SR architecture, the mapping is just the identity function: the global IPv6 address of the loopback interface of a node correspond to the SID of the node itself. For the MPLS SR architecture, a deterministic mapping from the IP address of the loopback interface used as router ID into a subset of the MPLS label space is needed. Assuming that the IP addresses of the loopback interfaces of the nodes belong to a contiguous range of IP addresses, this mapping is typically as simple as extracting the $N$ rightmost bits of the IP address and then offsetting the resulting value in a specifically allocated portion of the MPLS label space (e.g., $N=16$, if we want to allow for 65536 different nodes in the IGP routing domain, while the whole available MPLS label space is of 20 bits).

Mapping arbitrary prefixes into SIDs with an automatic procedure is not so easy. Therefore, we simply consider not to use Prefix-SIDs in our simplified architecture. We will show that we do not lose too much functionality with this choice. On the other hand, we cannot get rid of Adj-SIDs, for the reasons explained in the previous section. Hence, in order to avoid the use of local Adj-SIDs, we propose the introduction of a new type of global segment called *direct-link* segment. A *direct-link* segment identifies a target destination node to be reached (this is similar to a Node segment). If a node has a direct link toward the destination node, the *direct-link* segment forces the node to use the direct link rather than the shortest path dictated by the routing protocol. Conversely, if a node does not have a direct link toward the target node, it will process the segment in the same way it processes a Node segment toward the same destination node. We define a class of SIDs with global significance and corresponding to the *direct-link* segments, called *direct-link SID* or *DL-SID*. A *DL-SID* needs to identify the target node, like the Node-SID, and to carry further information that identifies it as *direct-link* SID. When using MPLS as Data Plane, the *DL-SID* can be obtained by the Node-SID adding a bit to distinguish between *DL-SID* and Node-SID. When using IPv6 as Data Plane, *DL-SID*s are IPv6 addresses globally valid in the network domain. They need to be derived in a deterministic way from the loopback interfaces addresses used as Node-SIDs. As an example, Node-SIDs can be restricted to have an odd numbered Device address part of the IPv6 address, so that the *DL-SID*s will be even numbered, obtained by adding one to the Device address part of the IPv6 address of the localhost interface.

A limitation of the proposed *DL-SID* approach is that it does not allow the handling of multiple parallel links between two routers at layer 3, i.e. with different IP addresses. If present, such *multi-links* must be handled at layer 2 and seen at IP level as a single link. Having multiple parallel links bonded at layer 2 is anyway a typical solution for operators, so we believe that it is not a critical limitation.

There are advantages in using the automatically generated global *DL-SID*s rather than the local Adj-SIDs or the global Adj-SID. Consider the *strict source routing* case, that is enforcing a path through a set of links. Using local Adj-SIDs, the segment list will have a length equal to the double of the number of links. In fact, for each link to be crossed, first the source node needs to be addressed, then the local segment will indicate the outgoing link. Using global Adj-SIDs, the list will be equal to the number of links, but the global Adj-SIDs needs to be advertised and one entry for each advertised Adj-SIDs needs to be added in the routing state of all nodes. In PMSR, with automatically generated global *DL-SID*s the length of a segment list to enforce a path through a set of links also equals the number of links (like with the global Adj-SID), but there is no need to advertise SIDs, and in each node it is only needed to add an additional entry for each node instead that for each link.

Let $\eta$ be the number of nodes and $k$ be the number of unidirectional links; in the worst case, $k=O(\eta^2)$.

| Traditional Segment Routing | | PMSR |
|---|---|---|
| Need to configure nodes with SIDs | | |
| Yes, $O(\eta)$ nodes | | No |
| **Local Adj-SID** | **Global Adj-SID** | **Autom. generated global DL-SIDs** |
| Need to advertise SIDs | | |
| Yes | Yes | No |
| Routing state | | |
| $O(\eta)$ | $O(\eta+k)$, $O(\eta+\eta^2)$ | $O(2\eta)$ |
| SR path length for a path of $\lambda$ links | | |
| L1 ≤ 2λ | L2 ≤ λ | L3 ≤ L1 |

Table 1 – Traditional SR vs. PMSR

Note that, in case of *strict source routing*, a list of local Adj-SIDs corresponding to the number of the links would actually be enough. However, we do not consider this solution for two reasons. First, because it is critical in case of failures of nodes/links in the path: intermediate nodes cannot reroute the packets and protection should be enforced edge-to-edge. Second, we are interested to the case of *loose source routing* (i.e. the segment list only includes a subset of the nodes in the path), because we want to use a small number of segments to create paths in SR. If *loose source routing* is used, we will see that in some cases a *DL-SID* could be not enough to uniquely identify a specific path and a couple of Node-SID + *DL-SID* will be needed. Table 1 summarizes the comparison between using PMSR (with global *DL-SIDs*) and the traditional SR with Local and Global Adj-SIDs.

A. *Node tables update procedures*

The SR-capable forwarding nodes need to populate their forwarding tables with entries related to the SIDs. In the traditional SR architecture, besides the Control Plane enhancements to distribute the SIDs, proper mechanisms to insert/update the forwarding table entries are needed. As an example, when receiving an announcement for a prefix-SID, the node will add an entry for the SID. If the forwarding architecture of the node allows it, the entry will be a "pointer" to the existing routing entry for the prefix. In this way, the routing toward the prefix can change, but the entry for the SID does not need to be updated. If it is not possible to add the "pointer", the entry for the SID needs to explicitly specify the next hop/outgoing interface and, in this case, it needs to be updated later if the routing towards the prefix will change. In the MPLS-based architecture, the SID is a MPLS label, therefore an entry will be added to the label forwarding tables, either specifying a logical link between the label and the IP forwarding information of the prefix or providing the indication of the next hop/output interface.

In the proposed PMSR architecture, the procedures for populating the forwarding tables are very simple and they do not rely on the processing of extensions to routing protocols. The entries for Node-SIDs and *DL-SIDs* are added following the routing information for the loopback addresses of the network nodes in the domain. For each entry related to a node loopback address there will be one entry for the corresponding Node-SID and one for the corresponding *DL-SID*. As discussed above, if it is possible to have a "pointer" to the routing entry for the remote loopback address, the entry will not need to be updated later on, otherwise the entry will contain the next hop/outgoing interface towards the remote loopback address and it will need to be updated if the routing changes.

For each remote loopback address to be added, the following steps are needed: 1) evaluate the Node-SID and the *DL-SID* for the remote node IP loopback address; 2) add(update) the entry for the Node-SID, pointing to the routing entry or extracting the next hop/outgoing interface from the routing entry; 3) if the node does not have a direct link toward the remote node, add(update) the entry for the *DL-SID* in the same way as described in step 2) for the Node-SID; if the node has a direct link toward the remote node, add the entry for the *DL-SID* pointing to the direct outgoing link, irrespective of the routing information. The evaluation of the Node-SID and *DL-SID* for the remote node depends on the Data Plane technology: for MPLS a label will be evaluated, for IPv6 an IP address will be considered. The addition of the entries will be performed in the label forwarding tables for MPLS or in the IP forwarding tables for IPv6.

### B. Analysis of the use cases

In the following table we report which use cases, among those presented in [4] [5] [6], are supported by the PMSR architecture. In general, all use cases which do not require the Prefix segments are well supported.

| Use case | Support |
|---|---|
| IGP-based MPLS Tunneling [4] [5] | OK |
| Fast Reroute [4] [5] [6] (Management free local protection and Managed local protection) | OK |
| Path Protection [6] | OK |
| Load balancing among non-parallel links [5] | NO[1] |
| Capacity Planning Process [4] [5] | OK |
| SDN/SR use case [4] [5] | OK |
| Service Chaining [5] | Easy[2] |
| OAM [5] | OK |
| Interoperability with non-Spring nodes [4] | OK |
| Disjointness in dual-plane networks [4] [5] | OK[3] |
| CoS-based Traffic Engineering [5] | OK[3] |
| Egress Peering Traffic Engineering [4] [5] | [4] |
| Distributed CSPF-based Traffic Engineering [5] | OK |
| Deterministic non-ECMP Path [5] | OK |

[1] This use case requires the advertising of a special adjacency segment that represents multiple outgoing links. In PMSR, this could be solved with workarounds based on SDN approach.
[2] In order to support Service Chaining new locally scoped SIDs have to be introduced. This can be easily introduced in PMSR with a SDN approach that avoids the need for advertising the local SIDs using routing protocols.
[3] These use cases include Anycast segments. There is no substantial difference between these segments and the Node segments used in PMSR.
[4] This use case includes BGP peering segments, which are local segments distributed using BGP protocol. PMSR behaves exactly like traditional SR here: it can support this use case, but it does not avoid the need of distributing information with BGP.

Table 2 – Use cases

From the analysis of the use cases, we realized that most of the use cases only require the Node-SIDs. In these cases, PMSR directly applies bringing the clear advantage of automatic generation of SIDs with no need to enhance routing protocols. Some TE related use cases require the use of Adj-SID, which in PMSR are mapped into *DL-SID*s. Therefore, in the rest of the paper we identify a TE use case that requires Adj-SID in the traditional SR architecture, and analyze the implications of using *DL-SID*s in the PMSR architecture.

### IV. TRAFFIC ENGINEERING USE CASE

The flow assignment problem consists in assigning a path to a set of flows. In a Segment Routing context, two types of flow assignment problems can be addressed: 1) ECMP-aware SR path assignment; 2) traditional *hop-by-hop* path assignment. The former is based on the identification of a set of nodes to be crossed, assuming that the flow will be evenly spread between the set of equal-cost paths towards the next segment by each node in the path. Under this assumption of even load distribution, it is still possible to evaluate the resulting load on each link, given the bandwidth requirement of the flow and the routing tables of all nodes. The traditional *hop-by-hop* path assignment does not rely on load sharing performed by nodes, because a single path for a flow is deterministically assigned. The resulting load on each crossed link simply corresponds to the bandwidth requirement of the flow.

In general, the capacity of exploiting Traffic Engineering based on the ECMP-aware path assignment is one key advantage of Segment Routing, with respect to traditional TE architectures (e.g., based on MPLS) that are only capable of working with *hop-by-hop* paths. Anyway, there can be use cases that advocate the use of deterministic *hop-by-hop* paths. As an example scenario, consider flows corresponding to single TCP connections. The ECMP output link selection is performed hashing the TCP ports and it will deterministically select a single output link for each crossed node. The assumption of even load sharing across the different ECMP paths is not verified in this case, leading to a mismatch between the planned and the actual resource allocation. Another scenario that calls for deterministic *hop-by-hop* paths is that of network topologies which do not present enough multiple equal-cost paths among source and destination nodes. Considering these scenarios, we think that the traditional *hop-by-hop* flow assignment problem represents a Traffic Engineering use case worth considering in a SR architecture. In section IV.A we present the TE problem along with a known formulation and heuristic resolution taken from the literature. In section IV.B we introduce the issue of mapping the *hop-by-hop* path into a list of segments. Section V describes the proposed SR path allocation mechanism and proves its optimality. In section VI some evaluation results are discussed. The analysis, proposals and results presented in these sections are not limited to PMSR but are fully applicable to traditional SR architecture.

### A. Hop-by-hop flow assignment: problem definition and heuristic resolution

Let $F$ be a set of unidirectional flows $f_i(s_i,d_i,r_i)$, where $s$ is the source node, $d$ the destination node and $r$ the nominal bandwidth requirement (b/s); let $T(N,E)$ be a directed graph representing the topology, $N$ is the set of nodes and $E$ is the set of directed edges. An edge $e_j$ can be represented as $e_j(u_j,v_j,c_j)$, where $u_j$ is the source node, $v_j$ the destination node and $c_j$ the edge capacity (b/s). An edge can also be denoted simply as $e(u,v)$, where $u$ is the source and $v$ the destination. Each flow $f_i$ needs to be mapped into an *hop-by-hop* path $P_i$ that can be represented as the set of intermediate nodes from source $s$ to destination $d$ (denoted as $Pn_i$), or equivalently by the set of links ($Pe_i$):

$Pn_i = \{ n_{i0}=s, n_{i1}, n_{i2}, .. , n_{iN-1}, n_{iN}=d \}$
$Pe_i = \{ e_{i1}, e_{i2}, .. , e_{iN-1}, e_{iN} \}$ where
$e_{i1}=e(s, n_{i1}), e_{i2}=e(n_{i1}, n_{i2}),… e_{iN}=e(n_{iN-1}, d)$

The traditional *hop-by-hop* path assignment consists in finding an "optimal" set of paths $\{P_i\}$, i.e. a set chosen according to an optimality criterion. Let us define the flow mapping variables $a_{ij}$, which tells if flow $f_i$ is mapped over link $e_j$: $a_{ij}=1$ if $e_j \in Pe_i$, $a_{ij}=0$ if $e_j \notin Pe_i$. In our formulation we also include a feasibility check: the sum of the nominal flow rates of the flows crossing a link needs to be smaller than the link capacity. In symbols:

$\forall$ link $j$: $\sum_i a_{ij} \cdot r_j < c_j$

For our experiments we reused (with few changes) the definition of the flow assignment problem and the heuristic for its resolution originally proposed in [10] and [11] (further details are given in [9]). The problem formulation is very effective in equalizing the load of the links in the network and avoiding critical bottleneck. In addition, the heuristic provides a good trade-off between computation time and optimality of results. Anyway, in this paper we are not interested in the quality of the heuristic or in the details of the TE optimization. We just take as input the set of *hop-by-hop* path allocated by the TE algorithm and consider their mapping into SR paths.

### B. Mapping hop-by-hop paths into SR paths

A Segment Routing path (SR path) will be denoted as $S_i$ and represented as a sequence of SIDs $Sn_i$:

$Sn_i = \{n_{i0}=s, n_{i1}, n_{i2}, .. , n_{iN-1}, n_{iN}=d\}$

In PMSR, each SID can be a Node-SID or a *DL-SID* (in the traditional SR architecture, a SID can also be a local or global Adj-SID, corresponding to an outgoing adjacency). A Node-SID is simply represented by the node name $n_1$, while the corresponding *DL-SID* is represented as $n_1^*$. In both cases, the SID corresponds to a node that needs to be crossed before reaching the destination node.

Two consecutive nodes in a SR path $Sn_i$ do not need to be adjacent as it is for $Pn_i$. When two consecutive nodes are not adjacent, the links that will be crossed depend on the underlying IP routing. If all the shortest paths from a given node toward the next node in the SR path insist on the same output link, then the output link is univocally determined. If there are multiple shortest paths and they insist on different output links, then the output link is not univocally determined. In this case, two options are possible, depending on the configuration of the router. If ECMP is enabled, all the "candidate" output links that are part of a shortest path towards the next node in the SR path are considered (typically they are selected based on a hash function over the port numbers of the transport protocol, in order to balance the traffic). If ECMP is not enabled, one of the candidate output links is arbitrarily selected by the node. In both cases, such type of segment is not applicable to the classical TE approach, in which the network operator wants to deterministically route a flow over a given path.

A SR path is congruent to a *hop-by-hop* path if the route enforced by the SR path is deterministically equivalent to the one enforced by the *hop-by-hop* path. To provide examples of *hop-by-hop* paths, of congruent SR paths, and of the use of *DL-SIDs*, let us consider the network topology depicted in Figure 1 and the two *hop-by-hop* paths $P_1$ and $P_2$ that are represented using $Pn$ notation as:

$Pn_1 = \{ n_1, n_3, n_5, n_7 \}$; $Pn_2 = \{ n_1, n_2, n_3, n_4, n_5, n_6, n_7 \}$

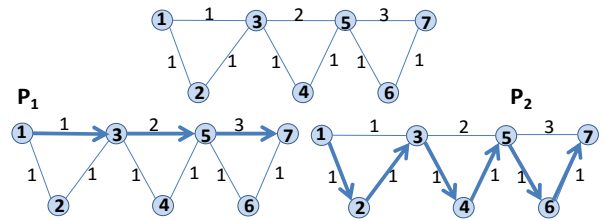

Figure 1 – A network topology and two hop-by-hop paths

The only SR path congruent to the *hop-by-hop* path $P_1$ is

$Sn_1 = \{ n_1, n_3, n_5^*, n_7^* \}$

in which three segments are needed, and the *direct-link* segment IDs $n_5^*$ and $n_7^*$ are respectively used to select the links 3→5 and 5→7.

There are multiple SR paths that are congruent to the *hop-by-hop* path $P_2$; a subset of them is listed hereafter (they only contain Node SIDs):

$Sn_{2-a} = \{ n_1, n_2, n_4, n_7 \}$
$Sn_{2-b} = \{ n_1, n_2, n_3, n_4, n_7 \}$

$Sn_{2-c} = \{ n_1, n_2, n_3, n_4, n_5, n_6, n_7 \}$

Among them, $Sn_{2-a}$ is the *optimal* SR path, in the sense that it has the minimum number of segments.

## V. OPTIMAL SR ASSIGNMENT PROCEDURE

In the SR assignment problem, given a *hop-by-hop* path $P$, we want to find a congruent SR path $S$ composed of the minimum number of segments. In this section we propose an efficient algorithm for the SR assignment, both for traditional SR and for the proposed PMSR. We prove that the algorithm finds the optimal solution, i.e. the shortest list of SIDs that allows the packets to follow the assigned *hop-by-hop* path, according to the default IP routing tables of the nodes. Let us define the following notation.

- *f*: a single traffic flow from node *s* to node *d*, characterized by its *hop-by-hop* path *Pn*:
  $Pn = \{ n_0=s, n_1, n_2, .., n_{N-1}, n_N=d \}$;
- *tep(x,y)*: portion of the *hop-by-hop* path starting from node *x* and ending with node *y*. As particular case, *tep(s,d)* is the complete *hop-by-hop* path from *s* to *d*;
- *SPN(x,y)*: the number of equal-cost shortest paths from *x* to *y*, based on the current routing tables that are considered to be already set-up by a link-state routing protocol (e.g. OSPF), using Shortest Path First algorithm;
- *sp(x,y)*: the set of the shortest paths from *x* to *y*; if $SPN(x,y) \equiv 1$, it is the shortest path from *x* to *y*;
- *prec(p,x)*: the preceding node of *x* along a path *p*;
- *succ(p,x)*: the succeeding node of *x* along a path *p*;
- *srp*: the SR path containing the list of assigned SIDs;
- $sp^*(x,y^*)$: the set of *direct-links biased* shortest paths from *x* to $y^*$; a *direct-links biased* shortest path is built heading from *x* to *y* on a shortest path, unless there is a *direct link* from an intermediate node to *y*, which is always followed;
- $SPN^*(x,y^*)$: number of *direct-links biased* shortest paths $sp^*(x,y^*)$.

A pseudo-code representation of the *SR assignment algorithm* for the traditional SR architecture is reported in Figure 2 (*T_SRP* stands for Traditional SR Path). The algorithm takes as input the topology and the assigned *hop-by-hop* path, and returns as output a congruent "optimal" SR path. At each step, a *hop-by-hop* sub-path between two nodes *x* and *y* is compared with the shortest path between the same pair of nodes. At the beginning *x=s* and *y=d*. If there is only one shortest path and it matches the *hop-by-hop* sub-path, *y* is added to the SR path. Otherwise (i.e., if there is more than one shortest path or the shortest path does not match the *hop-by-hop* sub-path), if the sub-path *tep(x,y)* between *x* and *y* is just one link, then it means that there is a direct link between *x* and *y* different from the shortest path; in this case the Adj-SID corresponding to the link *e(x,y)* is added to the SR path. If *tep(x,y)* is more than one link, the procedure repeats with *y* set to the node that precedes the old *y*. If a segment has been added, it is checked if $y \equiv d$, in which case the procedure ends and the SR path is returned; otherwise, if $y \neq d$, the algorithm considers the remaining part of the path, from *y* to *d*. For each direct link different from the shortest path, this algorithm will add two segments in the SR path: the preceding node and the Adj-SID representing the outgoing link.

```
function T_SRP: (tep(s, d)) → srp
  x = s; y = d;  srp = { }
START:
  p = tep(x, y);
  // check if the sub-path p is the only shortest path
  if ((SPN(x, y) == 1) AND (sp(x, y) == p)) then
      ADD y to srp; goto ADDED:
  else
      // check if the sub-path p is just one link
      if (prec(p, y) == x) then
          ADD Adj-SID of e(x,y) to srp; goto ADDED:
      else
          // no segment added, try with a shorter path
          // (from x to the node that precedes y)
          y = prec(p, y); goto START:
ADDED:
  if (y != d) then
      // consider the remaining part of the path
      x = y ;y = d; goto START:
  return srp;
```
Figure 2 – Pseudo-code of SR path assignment for traditional SR

```
function DL_SRP: srp → dlsrp
  dlsrp = { }
  for (i = 0; i < srp.length; i++)
      if (srp[i] is an Adj-SID) then
          d = destination of srp[i];
          ADD d* to dlsrp;
      else
          if (srp[i+1] is not an Adj-SID) then
              ADD srp[i] to dlsrp;
          else
              if (SPN*(srp[i-1],srp[i+1]) > 1 OR
                  sp*(srp[i-1],srp[i+1]*) != tep(srp[i-1],srp[i+1]))
              then
                  ADD srp[i] to dlsrp;
  return dlsrp;
```
Figure 3 – Replacement of adjacency SIDs with *direct-link* SID

The *DL_SRP* algorithm reported in Figure 3 takes as input the SR path (that includes Adj-SIDs) computed by *T_SRP* and returns, as output, a SR path that includes only Node-SIDs and *DL-SIDs*. When possible, it replaces a couple of Node-SID + Adj-SID with a single *DL-SID*. When a single *DL-SID* is not enough to enforce the required *hop-by-hop* path, the algorithm will leave a couple Node-SID + *DL-SID*. The algorithm inspects step-by-step the SR path and replaces any Adj-SID with the corresponding *DL-SID*. The Node-SID that precedes the Adj-SID is kept only when required, that is when there is more than one *direct-links biased* shortest path from the node that precedes the current Node-SID and the successive *DL-SID*, or if such a *direct-links biased* shortest path differs from the *hop-by-hop* path.

### A. Optimality of the SR path assignment

In order to demonstrate the optimality of the SR path assignment, we need the following Lemmas.

*Lemma 1*: if there is a unique shortest path from *s* to *d*, then there is a unique shortest path from *s* towards all intermediate links in the path from *s* to *d* (it can be easily proven by contradiction).

*Lemma 2*: if it does not exist a unique shortest path from $y$ to $d$, then it does not exist a unique shortest path from a node $x$ to $d$ that passes through $y$ (it can be easily proven by contradiction).

We start by focusing on the *T_SRP* algorithm. Let us consider the *hop-by-hop* path $Pn = \{n_0=s, n_1, n_2, .. , n_{N-1}, n_N=d\}$. Assume that the directed edge from $n_{k-1}$ to $n_k$ is not the shortest path from $n_{k-1}$ to $n_k$ (or it is one of a set of equal-cost shortest paths), then an Adj-SID is needed to enforce the use of the link $e(n_{k-1},n_k)$. Under this hypothesis, starting from $s$ the *T_SRP* algorithm can find one or more segments up to $n_{k-1}$ (the last segment being $n_{k-1}$ itself), but then it will identify the link that requires the Adj-SID (the first check "if the sub-path $p$ is the only shortest path" fails and the second check "if the sub-path $p$ is just one link" is verified) and add it. This happens for all the links that are not the shortest path between their source and destination. In the end, the SR path will be composed at least by all the Adj-SIDs, needed in order to route the packets on links that are, by definition, off the shortest path dictated by the routing protocol. Each Adj-SID will be preceded in the SR path by the Node-SID of the node that originates the link that requires the Adj-SID. Now we need to demonstrate that the number of segments, selected by the algorithms in any portion of the *hop-by-hop* path that does not need to include Adj-SIDs, is the minimum possible. Assume from now on that we are in a portion of the *hop-by-hop* that does not need to include Adj-SIDs (i.e. all links correspond to the only shortest path between source and destination of the link). The *T_SRP* algorithm starts from the source $s$ and tries to find the longest portion of the *hop-by-hop* path $P=tep(s,d)$ that corresponds to a shortest path. If it arrives to the destination $d$, then the solution is optimal. If it stops at an intermediate node $x$, this means that $tep(s,x)$ is a unique shortest path, while $tep(s,succ(P,x))$ is not a unique shortest path. The algorithm tries to find segments from $x$ to $d$. If there is a unique shortest path from $x$ to $d$, then the algorithm has found a SR path with two segments: $\{s,x,d\}$. This is optimal, as a solution with one segment does not exist (we know that $tep(s,succ(P,x))$ is not a unique shortest path and, by Lemma 1, there cannot be a unique shortest path from $s$ to $d$). If the algorithm finds that an intermediate node $y$ is needed from $x$ to $d$, then we have a three segments solution: $\{s,x,y,d\}$, and we prove that we cannot find a two segments solution $\{s,z,d\}$ for any $z$ in $P$. In fact, the segment $z$ cannot be after $x$ by construction. It cannot be before $x$ because by Lemma 2 there cannot be a unique shortest path from $z$ to $d$ passing through $x$. This reasoning can be extended to any number of segments: each time that the algorithm introduces a segment, it is not possible to find a solution with a smaller number of segments.

It is easy to prove that the *DL_SRP* algorithm is optimal as well. In fact, it includes one *DL-SID* for each Adj-SID (they correspond to the minimum number of segments). In each portion of the path without Adj-SID, the algorithm verifies if it is possible to reduce the segments eliminating the last Node-SID and using only the *DL-SID*.

## VI. IMPLEMENTATION AND EVALUATION

The PMSR solution and TE algorithms have been implemented, further details (referring to a simpler, earlier version) are described in [8][9]. The source code is available at [13], including the Java implementation of the flow assignment and SR path assignment algorithms. A ready-to-go virtual machine is available ([12]).

Hereafter we report a simple experimental evaluation of the processing time of the proposed *DL-SID*-based *SR assignment* algorithm. We considered a relatively large scale topology (Figure 4) with 153 nodes and 354 unidirectional links, the "Colt Telecom" topology which is included in the Topology zoo dataset [14], assuming that all links have the same capacity. We generated a random set of traffic demands as follows. We randomly selected 40% of the nodes to be PE (e.g., ingress/egress), then we randomly selected 20% of the PE couples to be active source/destination of traffic flows. For each active couple of PEs, in each direction we have an average of 3.5 flows (the number of flows has a geometrical distribution) with the sum of the flow rates equals to 10% of the capacity of a link and the size of each flow that has a negative exponential distribution. With these parameters, we generated a list of 2460 flows along their bit rate. This demand largely overcomes the network capacity, so that only 940 flows can be allocated using the implemented algorithms. We selected only the accepted flows, obtaining a traffic demand that closely matches the full network capacity, being able to have *hop-by-hop* paths that diverge from the shortest path, but keeping the acceptance ratio of the flow close to 1.

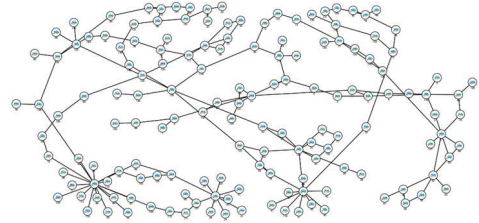

Figure 4 – Colt Telecom (08/2010) topology from Topology Zoo (each link in the picture corresponds to two unidirectional links)

Figure 5 reports the time spent for the computation of TE paths (flow assignment heuristic) and of SR paths (*SR assignment* algorithm). We use a PC with an Intel Core i7 2Ghz and 6GB RAM. Note that processing time of the flow assignment heuristic has a step-wise dependence on the number of iterations of the heuristic optimization cycle, which tends to increase with the number of flows. Therefore a set of seemingly parallel lines can be appreciated in the figure (each one corresponds to a given number of cycles). As it is possible to see from the figure, the processing time of the *SR assignment* algorithm is negligible with respect to the flow assignment heuristic. In the considered range (up to 900 admitted flows) it was possible to run both algorithms and allocate the flows in less than 8 seconds. This performance seems adequate for periodic (e.g., nightly) reallocation procedures that aim to evenly redistribute the load on the network links.

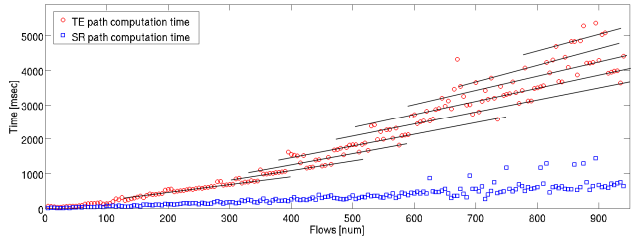

Figure 5 – Execution time of the algorithms

## VII. STATE OF THE ART AND RELATED PROJECTS

The Segment Routing architecture is being standardized within the IETF by the SPRING working group [17]. In sections I and II we have introduced SR technology and provided references to the active draft specifications.

SR-IPv6 [18] provides an Open Source implementation of IPv6 Data Plane for SR. Control Plane and Traffic Engineering aspects are not covered in [18].

The SPRING-OPEN project [19] is an ONOS [20] use case, which provides an SDN-based implementation of SR. Its architecture is based on a logically centralized Control Plane, built on top of ONOS, and it drastically eliminates the IP/MPLS Control Plane from the network. Compared to SPRING-OPEN, our solution still considers a traditional IP Control Plane (e.g., based on routing protocols like OSPF or IS-IS).

In both [21] and [22] the authors deal with SR-based ECMP-aware Traffic Engineering, proposing solutions for the optimal allocation of traffic demands using an ECMP-aware approach. Our TE problem is different, as we start from *hop-by-hop* paths and try to optimize their mapping into SR paths, keeping the constraint of the fixed routing over the given *hop-by-hop* path.

In [23] two SR testbeds are described, one based on a SDN scenario and another one based on a PCE scenario. Both testbeds share a common SR Path computation engine, that performs the *hop-by-hop* path computation and SR path assignment. The proposed SR path assignment algorithm provides the shortest segment list, but the solution only considers global Node-SID, therefore it cannot be applied to topologies with arbitrary IGP link costs. In [24] a rather general TE algorithm for SR is considered. It evaluates an optimal path for a flow, according to an IGP metric and taking into account bandwidth and delay constraints; then it minimizes (or enforces a bound on) the number of segments. It considers ECMP forwarding by default, but can also introduce constraints to support a deterministic *hop-by-hop* path. The solution is not able to support arbitrary *hop-by-hop* paths when arbitrary IGP link costs are used.

## VIII. CONCLUSIONS

In this paper we presented PMSR, a Segment Routing solution that does not require enhancements to routing protocols. PMSR is based on the use of global segment identifiers that can be automatically generated by nodes. We discussed the advantages of PMSR (in terms of simplification of management and reduction of node complexity) and advocated the suitability of PMSR to support the typical SR use cases. As the PMSR requires the introduction of *Direct Link* Segments to replace traditional SR Adjacency Segments, we considered a Traffic Engineering use case that requires the Adjacency Segments. We proposed an algorithm for the SR path allocation, useful for both traditional SR with Adjacency Segments and for PMSR with *direct-link* Segments. We proved that it is optimal in terms of the number of allocated segments and empirically verified that the execution time is small compared with the TE heuristic preliminarily needed to allocate the *hop-by-hop* path.

## ACKNOWLEDGMENTS

This work builds on the results of DREAMER project, partly funded by the EU as one of the beneficiary projects of the GÉANT Open Call research initiative.